# MRI/TRUS data fusion for brachytherapy

V. Daanen[1,2], J. Gastaldo[3], J.Y. Giraud[3], P. Fourneret[3], J.L.Descotes[4], M. Bolla[3], D. Collomb[5], J. Troccaz[1]

[1]TIMC Laboratory, Grenoble, France
[2]Medical informatics Dept, Grenoble University Hospital, France
[3]Radiation Oncology Dept, Grenoble University Hospital, France
[4]Urology Dept, Grenoble University Hospital, France
[5]Radiology Dept, Grenoble University Hospital, France

Author for correspondence: Jocelyne Troccaz, TIMC, IN3S, Faculté de Médecine, Domaine de la Merci, 38706 La Tronche cedex, France. Email : jocelyne.troccaz@imag.fr

**Abstract**
*Background*: Prostate brachytherapy consists in placing radioactive seeds for tumour destruction under transrectal ultrasound imaging (TRUS) control. It requires prostate delineation from the images for dose planning. Because ultrasound imaging is patient and operator-dependant, we have proposed to fuse MRI data to TRUS data to make image processing more reliable. Technical accuracy of this approach has already been evaluated.
*Methods*: We present work in progress concerning the evaluation of the approach from the dosimetry viewpoint. The objective is to determine which impact this system may have on the treatment of the patient. Dose planning is performed from initial TRUS prostate contours and evaluated on contours modified by data fusion.
*Results*: For the 8 patients included, we demonstrate that TRUS prostate volume is most often underestimated and that dose is overestimated in a correlated way. However, dose constraints are still verified for those 8 patients.
*Conclusions*: This confirms our initial hypothesis.

## 1. Introduction

Prostate cancer is one of the most common malignancy among men. An estimated 543000 cases and 204000 deaths were attributed to prostate cancer in 2000 [1]. Its detection is based on digital rectal examination (DRE) and Prostate Specific Antigen (PSA) rating and is confirmed through the anatomo-pathologic analysis of biopsies. Several treatments may be proposed to the patient suffering from localized prostate cancer (T1c-T2b N0 M0, UICC 2002). One of them is prostate brachytherapy which consists in placing permanent radioactive seeds, most often $I^{125}$, inside the prostate thanks to needles passing through the perineum. Most often treatment planning and needle placement rely on the intensive use of transrectal ultrasound imaging (TRUS). Parallel transverse images are acquired with constant inter-slice distance thanks to the use of a stepper. Needle insertion through the perinea is made easier thanks to a grid rigidly connected to the TRUS probe (see figure 1).

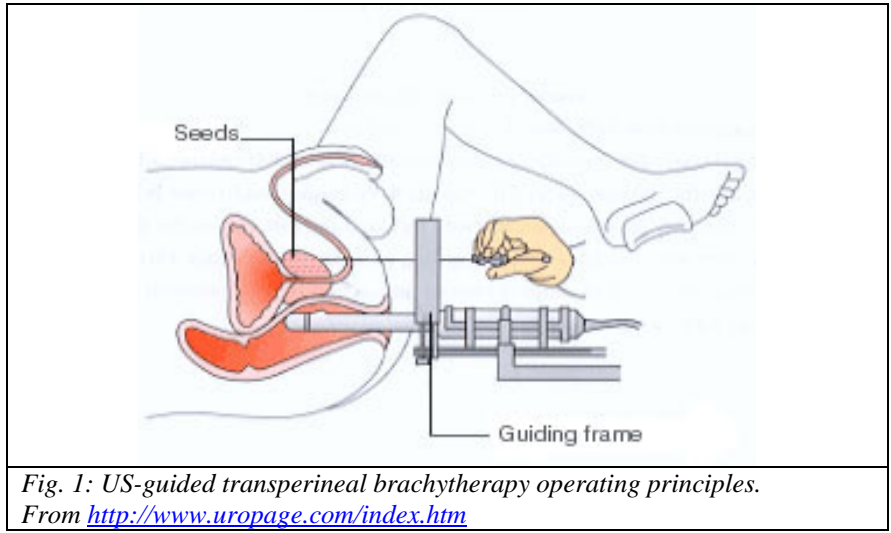
*Fig. 1: US-guided transperineal brachytherapy operating principles. From http://www.uropage.com/index.htm*

Prostate volume study is a critical step for treatment planning; it allows selecting precisely the number of seeds to be inserted and their position in the gland in order to obtain a predefined dose distribution (see section 3). Traditionally, this volume definition is manually performed by the clinician from TRUS images (either pre-operatively or intra-operatively). But it is well known that ultrasonic image quality is highly variable from patient to patient and that organ delineation varies from user to user and for a same user, from session to session. In this context, Magnetic Resonance Imaging (MRI) has a great potential. In very few centres, radioactive seeds are implanted inside an open interventional MR machine [2,3]. Robots may also be used in conjunction to MRI imaging for improved accuracy and increased possible trajectories [4]. Often such robots or systems are jointly designed for biopsy and brachytherapy since both gestures basically involve needle insertion into the prostate under TRUS control.

In this project, we chose to *bring the MR virtually in the Operating Room* (OR) *by TRUS/MRI data fusion* in order to help the clinician for treatment planning without requiring the presence of both the patient and the implantation team in the MRI room. A similar idea was used by [5] in the context of TRUS-guided prostate biopsies. However, where [5] uses rigid registration of 6 anatomical landmarks, the present work makes use of a combined rigid and elastic surface registration.

Having demonstrated the accuracy of data fusion (see [6]), a study has been undertaken for evaluating the impact of MRI/TRUS data fusion on dose distribution. This on-going work is described in the paper. Section 2 presents the method and summarizes preliminary results. Section 3 introduces dose constraints to be satisfied. Section 4 describes the evaluation protocol. Results are given in section 5 and discussed in section 6.

## 2. MRI/US data fusion: principles and previous results

The developed system is based on combined rigid and elastic surface registration. Pre-operatively, a MRI acquisition of the patient prostate is performed; three orthogonal T2 TSE volumes are obtained using an endorectal coil and the prostate contours are segmented jointly on the three volumes. This gives a set of points $S^{MRI}$ defined in a single referential $R_{MRI}$. Intra-operatively, TRUS axial slices are collected and manually segmented by the urologist as in the conventional procedure and this results in a sparser intra-operative 3D prostate model $S^{TRUS}$ defined in a referential $R_{TRUS}$; as mentioned above, the axial contours are necessary for dose simulation and planning. Pseudo-sagital slices may also be acquired and segmented and then merged to the data resulting from the axial acquisition (cf. figure 2). The objective of registration is to determine the transform between $R_{TRUS}$ and $R_{MRI}$ allowing the optimal superimposition of $S^{TRUS}$ and $S^{MRI}$. Registration determines prostate motion between the two acquisitions as well as prostate deformations. A pre-registration consisting in superimposing the centres of gravity of $S^{TRUS}$ and $S^{MRI}$ initializes the unknown transform. From this initial estimate, rotation and translation parameters between data sets as well as local deformations or distortions are determined. The method is derived from the octree-spline elastic registration described in [7]. It makes use of an adaptive, hierarchical and regularized free-form deformation of one volume to the other coordinate system. The optimization stage is based on the Levenberg-Marquadt minimization procedure. The result is a 3-D function *f* transforming any ultrasound data point in $R_{TRUS}$ to the corresponding MRI point in $R_{MRI}$.

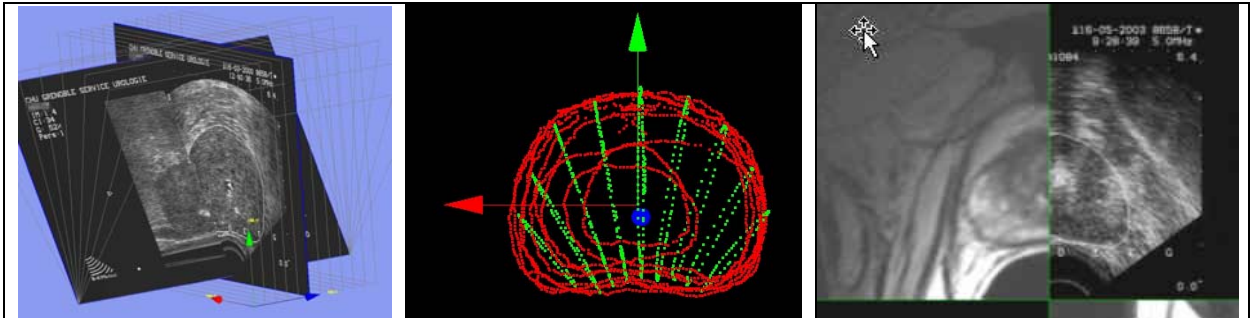

*Fig. 2.*
*MRI/TRUS fusion for brachytherapy: (left) two of the axial and pseudo-sagital TRUS images (middle) 3D TRUS reconstruction (right) composite image produced after MRI/TRUS elastic registration: for each axial TRUS image the corresponding MRI data are computed – the green cursor enable motion in the modalities.*

After MRI/TRUS surface-based elastic registration, for each initial TRUS image, a new composite image divided in 4 quadrants (2 TRUS quadrants and 2 MRI ones) is computed (see figure 2). The pixel intensity I at coordinates (u,v) of the composite image associated to the $n^{th}$ TRUS image is obtained as follows:

- I (u, v, n) = TRUS (u, v, n) in the TRUS quadrant where TRUS (u, v, n) is the pixel intensity at coordinates (u, v) in the $n^{th}$ TRUS image
- I (u, v, n) = MRI (x, y, z) where (x, y, z) = *f* (u, v, n) in the MRI quadrants; since x, y and z have real values MRI (x, y, z) is obtained by interpolating neighbour intensity values in the MRI volume.

The user can interactively modify the composite image by moving the cross cursor dividing this image; this enables exploring the prostate jointly in the MRI and TRUS data. This may result in modifying locally the TRUS segmentations of the axial slices when MRI data make prostate delineation easier. In the following prostate contours obtained after MRI/TRUS data fusion are named "MRI+US contours" whilst original contours are named "US contours". More generally, all data or measurements obtained after MRI/TRUS fusion are named "MRI+US xxx" where measurements from US contours are named "US xxx".

[6] presents a technical evaluation of the system on phantom and patient data. Both rigid and elastic registrations were evaluated; elastic registration which demonstrated higher accuracy is briefly reported below (see table 1). It was measured both in terms of residual distance between matched surfaces and using TRE (Target Registration Error) evaluated from the urethra which does not participate to matched data. Residual surface distance was measured

for 11 patients; TRE was evaluated for only 4 of them who were equipped with an endo-urethral probe making the urethra fully visible on TRUS and MRI modalities.

|          | Residual surface distance (mm) | TRE (mm)  |
|----------|-------------------------------|-----------|
| Phantom  | 1.07±0.41                     | 1.57±0.62 |
| Patients | 1.11±0.54                     | 2.07±1.57 |

*Table 1: Accuracy evaluation of elastic registration (see [6] for more details).*

Another aspect of the evaluation concerned prostate volume measurements from TRUS images with and without data fusion. Because the number of seeds depends on the prostate volume, modification in the latter may have a significant clinical impact. This preliminary study showed that apex and base US slices where often added to the segmented data when MRI information was fused to US data. For the 11 patients involved in the trial, the MRI+US gland volume was higher than the US volume; for elastic registration this volume increase was in average 15.86(±13.57)% of the US volume. Underestimating the prostate volume could result in delivering less radiation than expected to the target. Based on those preliminary observations, the project has been then to analyze MRI/TRUS fusion in prostate brachytherapy from the viewpoint of dose delivery to the organs.

## 3. Dose constraints

The objective is to evaluate the potential influence of TRUS/MRI registration on the treatment of the patient. In the Grenoble Hospital, the target dose for treatment of the prostate is 160Gy. The corresponding dose constraint is to deliver between 160Gy and 180Gy to 90% of the volume of the gland. Maximum dose constraints are also specified for the anatomical structures at risks: urethra and rectum. At least 30% of the urethra volume must receive less than 240Gy (150% of the target dose); a dose greater to 160Gy must not be delivered to more than 1.3cc of the rectum. Finally 90% of the rectum volume must receive less than 80Gy.

One common way to represent dose distribution is the so-called Dose Volume Histogram (DVH). The x coordinate represents a dose (percentage of total dose or absolute value in Gy) and the y coordinate denotes a volume (percentage of total volume or absolute value in cc). A point (x,y) on the curve associated to an organ or to an anatomical structure means for instance: y% of the organ volume receives at least x Gy. The constraints listed in the previous paragraph can be summarized in figure 3. D90 is the maximum dose received by 90% of the organ.

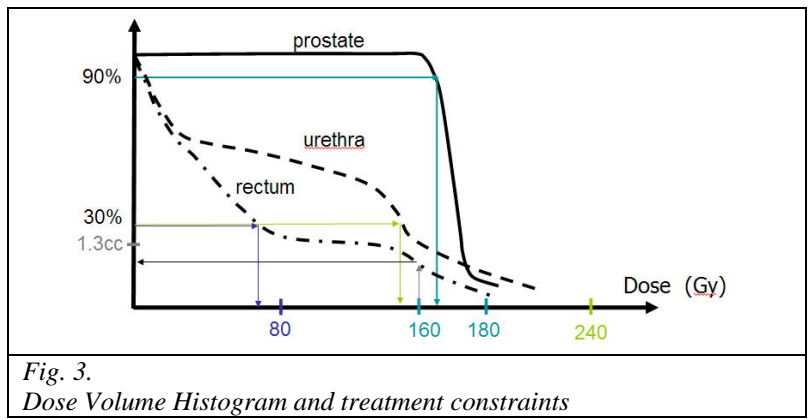

*Fig. 3.*
*Dose Volume Histogram and treatment constraints*

## 4. Proposed evaluation protocol

At this stage of the study the patient is treated in the conventional way. During the conventional procedure, data are collected and stored in the DICOM-RT format including US images, segmented contours and treatment plan. After patient treatment, those data are post-processed. MRI data are fused to TRUS information and segmented MRI contours are mapped onto TRUS images using the computed transfer function $f$. This gives a new set of contours in the US images; the dosimetric plan is evaluated for those two volumes: the US volume and the MRI+US volume.

Up to now 8 patients have been included in this protocol which has been approved by the hospital ethical committee; we plan to include a total of 16 patients. Different parameters have been examined and compared:
- prostate volumes,
- percentage of prostate volumes in the 140, 160 and 180Gy isodoses,
- values of D90 in the prostate.

## 5. Results

Table 2 presents the prostate volume analysis with and without data fusion for the 8 patients. As can be seen the MRI+US volume is greater than the US volume for 6 of the 8 patients. Most volume differences are in the apex and base regions of the prostate for which TRUS images are often difficult to interpret. For 3 of those 6 patients, the increase of volume exceeds 10% of the initial US volume. Since the number of patients is rather small, a non parametric test has been selected. A Wilcoxon test has been applied to those paired measurements which demonstrates that the two series of measurements are significantly different (p=0.05).

| Patient number | US volume (cc) | MRI+US volume (cc) | Diff = MRI+US volume - US volume (cc) | 100 Diff / US volume (%) |
|---|---|---|---|---|
| 1 | 24,12 | 23,05 | -1,07 | -4,43 |
| 2 | 21,22 | 22,91 | 1,684 | 7,93 |
| 3 | 44,21 | 50,53 | 6,32 | 14,29 |
| 4 | 26,74 | 28,72 | 1,98 | 7,40 |
| 5 | 29,52 | 32,66 | 3,14 | 10,63 |
| 6 | 31,09 | 35,05 | 3,96 | 12,73 |
| 7 | 35,93 | 38,91 | 2,98 | 8,29 |
| 8 | 36,57 | 35,64 | -0,93 | -2,54 |
| Mean value | 31,17 | 33,43 | 2,25 | 6,79 |
| Stand. dev. | 7,49 | 9,03 | 2,45 | 6,79 |

*Table 2: Prostate volume analysis*

The "isodose 160" encloses the volume of the prostate that receives at least 160Gy, the target dose. Table 3 presents the "isodose 160" analysis with and without data fusion. As can be seen, the MRI+US "isodose 160" includes less prostate volume for 7 of the 8 patients. However, all cases still verify the constraint of having 90% of the prostate receiving more than 160Gy. Patient 5 comes close to this inferior limit of the dose. A Wilcoxon test has been also applied to those paired measurements demonstrating that the two series of measurements are significantly different (p=0.05).

| Patient number | US "isodose 160" (% of prostate volume) | MRI+US "isodose 160" (%) | Diff = MRI+US isodose - US isodose (%) | 100 Diff / US isodose (%) |
|---|---|---|---|---|
| 1 | 96,67 | 96,76 | 0,09 | 0,09 |
| 2 | 98,08 | 93,7 | -4,38 | -4,46 |
| 3 | 98 | 91,34 | -6,66 | -6,79 |

| 4 | 98,39 | 95,49 | -2,9 | -2,94 |
| 5 | 97,41 | 90,5 | -6,91 | -7,09 |
| 6 | 99,43 | 93,18 | -6,25 | -6,28 |
| 7 | 97,82 | 92,41 | -5,41 | -5,53 |
| 8 | 98,63 | 97,49 | -1,14 | -1,15 |
| Mean value | *98,05* | *93,85* | *-4,19* | *-4,27* |
| Stand. dev. | *0,82* | *2,52* | *2,63* | *2,68* |

*Table 3: "Isodose 160" analysis*

As expected those dose modifications are correlated to volume ones (Spearman coefficient: 0.904 greater than Spearman threshold=0.88 for n=8, alpha risk=0.01); this is quite visible on figure 4.

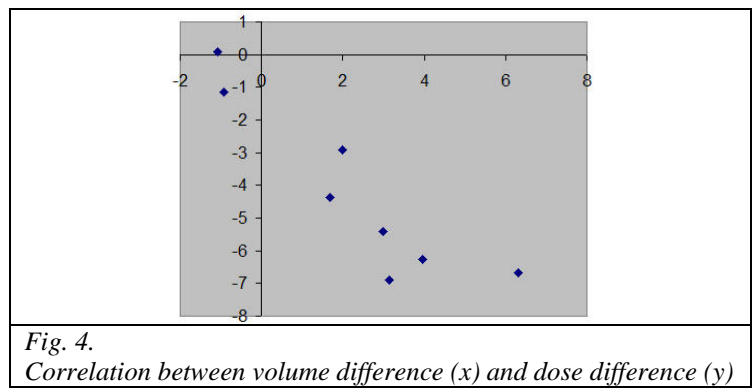

*Fig. 4.*
*Correlation between volume difference (x) and dose difference (y)*

D90 is the maximum dose received by 90% of the prostate. As mentioned below, this dose must be more than 160Gy and less than 180Gy. Table 4 presents a comparison of estimates of the D90 (since the exact value of D90 is difficult to get) with and without data fusion. The exact volume corresponding to the dose is mentioned near the dose value. As can be seen, the MRI+US D90 is smaller for 7 of the 8 patients. However, all cases still verify the constraints of having 160Gy<D90<180Gy. Patient 5 comes close to this inferior limit of the dose. A test of Wilcoxon has been applied to those paired measurements demonstrating that the two series of measurements are significantly different (p=0.05).

| Patient number | US D90 (vol) (Gy (%)) | MRI+US D90 (vol) (Gy (%)) | Diff = MRI+US D90 - US D90 (Gy) | 100 Diff / US D90 (%) |
|---|---|---|---|---|
| 1 | 175 (89,58) | 175 (90,09) | 0 | 0 |
| 2 | 175 (89,32) | 165 (92,25) | 10 | 5,71 |
| 3 | 175 (88,44) | 165 (89,89) | 10 | 5,71 |
| 4 | 179,2 (90,4) | 172,8 (90,44) | 6,4 | 3,57 |
| 5 | 172,8 (90,71) | 160 (90,5) | 12,8 | 7,40 |
| 6 | 174,4 (90,33) | 166,4 (90,45) | 8 | 4,58 |
| 7 | 174,4 (90,81) | 164,8 (90,27) | 9,6 | 5,50 |
| 8 | 174,4 (90,24) | 172,8 (90,07) | 1,6 | 0,91 |
| Mean value | *175,02 (89,97)* | *167,72 (90,49)* | *7,3* | *4,17* |
| Stand. dev. | *1,83 (0,8)* | *5,19 (0,74)* | *4,42* | *2,55* |

*Table 4: D90 analysis*

Figure 5 presents the comparative DVH of patient number 7.

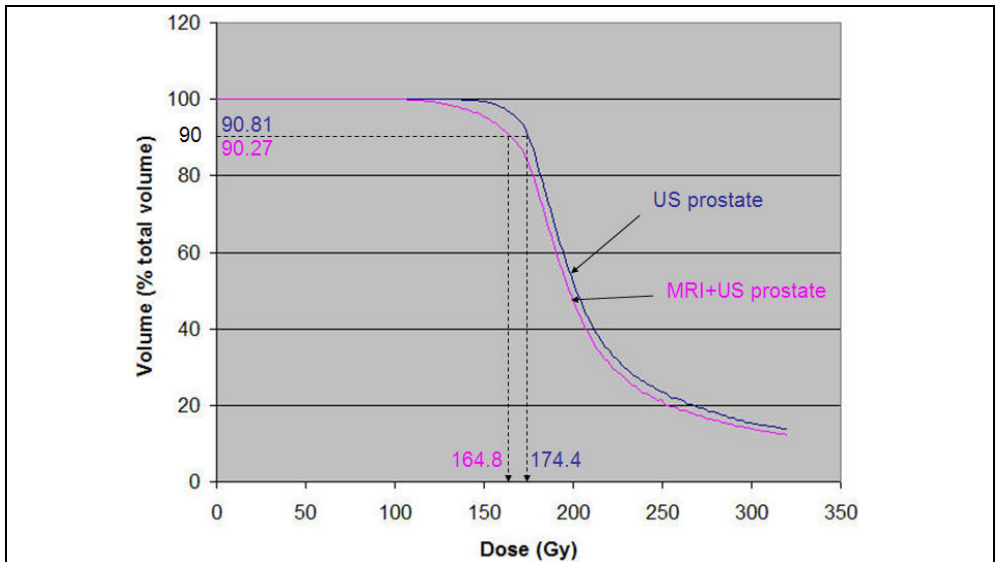
*Fig.5.*
*Comparative DVH for patient number 7*

For patient 3, for whom the volume difference is the greatest, figure 6 illustrates dosimetry on the apex (left and middle) and base (right) prostate TRUS images. MRI+US contours are presented in red; US contours are brown; yellow points represent planned seed positions; green and pink contours respectively correspond to 160Gy and 240Gy isodoses. The two apex images clearly show a potential underdosage of the prostate.

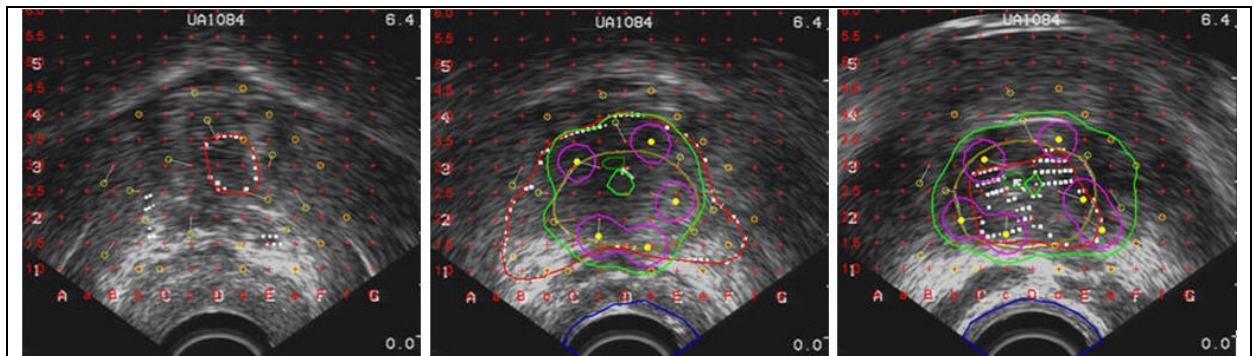
*Fig. 6.*
*Visualisation of MRI+US and US contours plus isodoses on apex (left; middle) and base (right) TRUS images.*

### 6. Discussion
Results presented below although still partial confirm the hypothesis that TRUS image delineation often underestimates prostate volume due to poor readability of apex and base slices and that it results in overestimating the dose delivered to the patient.

We observed that volume difference for those 8 patients was smaller than for the 11 patients having been included in the first evaluation. On average this volume difference was 15.86% for the 11 first patients instead of 6.79% for the 8 patients included in this work. Since the protocols were slightly different this does not allow a too direct comparison of numbers. Meanwhile, those differences may be explained partly by a side effect of this study having probably resulted in a learning curve associated to prostate TRUS delineation; indeed when the work on MRI/TRUS fusion began, the implantation team (urologist, radiation oncologist and radiophysicist) had still a rather low experience of prostate brachytherapy – implantations began in July 2001. We suppose that post-processing of data and comparison of US data to MRI ones increased the level of knowledge of the urologist who delineates the prostate on the TRUS images.

From the dosimetry point of view experiments confirm that the almost systematic underestimation of prostate volume results in a significant decrease of dose delivered to the prostate. However, prostate dose constraints are still verified on MRI+US contours for those 8 patients. This has to be confirmed on a larger number of patients since some preliminary evaluation exhibited cases where the D90 was substantially lower than expected (see [6]). This has also to be confirmed on organs at risk but this requires software modification which is in progress.

The overestimation of delivered dose has probably a limited effect in terms of treatment quality and therapeutic effects. Nevertheless the implantation team is very enthusiastic about using this system since they expect an improved comfort for image processing and an increased confidence in the planning.

One can see on figure 7 another use of the system where isodoses corresponding to planning from the TRUS images are mapped to the MRI data. The visual validation of those isodoses from MRI data is apparently much easier than the direct evaluation from TRUS images since anatomical substructures are more visible on MRI than on TRUS data. This point has not yet been tested extensively but clinicians and physicists reacted very positively to such a possibility. One prospective development of this project could be to perform the planning directly on MRI data and to transfer this planning to the OR thanks to MRI/TRUS data fusion.

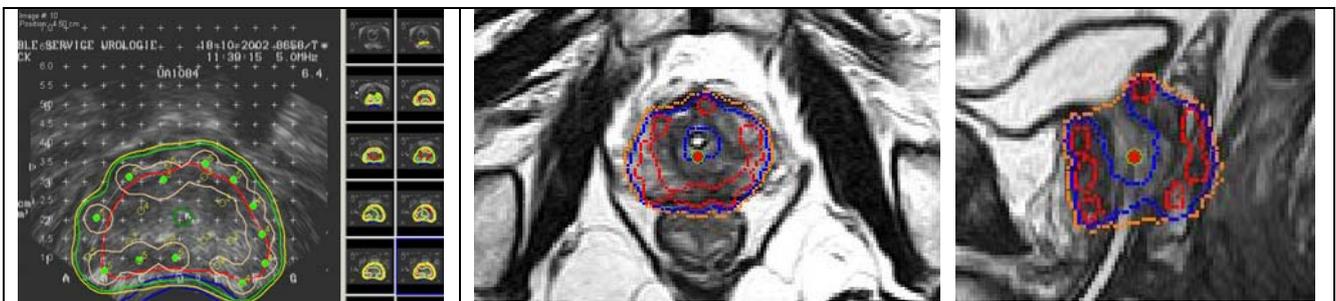

*Fig. 7.*
*Dose visual assessment (left) from TRUS image (middle and right) from MRI axial and sagital views after MRI/TRUS data fusion. For middle and right views: orange=150Gy isodose; blue=180Gy isodose; red=240Gy isodose.*

### 7. Conclusion
We have presented on-going work concerning the impact of data fusion on dose distribution for prostate brachytherapy. Dose is systematically overestimated due to underestimation of the TRUS prostate volume although the treatment still verifies dose constraints on the prostate for the 8 patients already included in this trial. Clinically using MRI/TRUS data fusion would result in a more reliable delineation of the prostate making real dose delivered to the prostate certainly closer to the planned dose. Further work remains concerning the evaluation of dose distribution on urethra and rectum. However, obtained results are already encouraging. From those preliminary results it is likely that MRI/TRUS data fusion could facilitate treatment planning and improve quality assessment.

**Acknowledgments**

This work is supported by the French Ministry of Health (Projet Hospitalier de Recherche Clinique PHRC National 2003 Prostate-Echo).

|  | Residual surface distance (mm) | TRE (mm) |
|---|---|---|
| Phantom | 1.07±0.41 | 1.57±0.62 |
| Patients | 1.11±0.54 | 2.07±1.57 |

*Table 1: Accuracy evaluation of elastic registration (see [6] for more details).*

| Patient number | US volume (cc) | MRI+US volume (cc) | Diff = MRI+US volume - US volume (cc) | 100 Diff / US volume (%) |
|---|---|---|---|---|
| 1 | 24,12 | 23,05 | -1,07 | -4,43 |
| 2 | 21,22 | 22,91 | 1,684 | 7,93 |
| 3 | 44,21 | 50,53 | 6,32 | 14,29 |
| 4 | 26,74 | 28,72 | 1,98 | 7,40 |
| 5 | 29,52 | 32,66 | 3,14 | 10,63 |
| 6 | 31,09 | 35,05 | 3,96 | 12,73 |
| 7 | 35,93 | 38,91 | 2,98 | 8,29 |
| 8 | 36,57 | 35,64 | -0,93 | -2,54 |
| Mean value | *31,17* | *33,43* | *2,25* | *6,79* |
| Stand. dev. | *7,49* | *9,03* | *2,45* | *6,79* |

*Table 2: Prostate volume analysis*

| Patient number | US "isodose 160" (% of prostate volume) | MRI+US "isodose 160" (%) | Diff = MRI+US isodose - US isodose (%) | 100 Diff / US isodose (%) |
|---|---|---|---|---|
| 1 | 96,67 | 96,76 | 0,09 | 0,09 |
| 2 | 98,08 | 93,7 | -4,38 | -4,46 |
| 3 | 98 | 91,34 | -6,66 | -6,79 |
| 4 | 98,39 | 95,49 | -2,9 | -2,94 |
| 5 | 97,41 | 90,5 | -6,91 | -7,09 |
| 6 | 99,43 | 93,18 | -6,25 | -6,28 |
| 7 | 97,82 | 92,41 | -5,41 | -5,53 |
| 8 | 98,63 | 97,49 | -1,14 | -1,15 |
| Mean value | *98,05* | *93,85* | *-4,19* | *-4,27* |
| Stand. dev. | *0,82* | *2,52* | *2,63* | *2,68* |

*Table 3: "Isodose 160" analysis*

| Patient number | US D90 (vol) (Gy (%)) | MRI+US D90 (vol) (Gy (%)) | Diff = MRI+US D90 - US D90 (Gy) | 100 Diff / US D90 (%) |
|---|---|---|---|---|
| 1 | 175 (89,58) | 175 (90,09) | 0 | 0 |
| 2 | 175 (89,32) | 165 (92,25) | 10 | 5,71 |
| 3 | 175 (88,44) | 165 (89,89) | 10 | 5,71 |
| 4 | 179,2 (90,4) | 172,8 (90,44) | 6,4 | 3,57 |
| 5 | 172,8 (90,71) | 160 (90,5) | 12,8 | 7,40 |
| 6 | 174,4 (90,33) | 166,4 (90,45) | 8 | 4,58 |
| 7 | 174,4 (90,81) | 164,8 (90,27) | 9,6 | 5,50 |
| 8 | 174,4 (90,24) | 172,8 (90,07) | 1,6 | 0,91 |
| Mean value | *175,02 (89,97)* | *167,72 (90,49)* | *7,3* | *4,17* |
| Stand. dev. | *1,83 (0,8)* | *5,19 (0,74)* | *4,42* | *2,55* |

*Table 4: D90 analysis*